\documentclass[aps,twocolumn,pra,superscriptaddress,tightenlines]{revtex4-1}
\usepackage{amsfonts}
\usepackage{graphicx}
\usepackage{hyperref}
\usepackage{amsmath}
\usepackage{amssymb}
\usepackage[usenames, dvipsnames]{color}
\usepackage{subfigure}
\usepackage{lipsum}

\begin{document}

\title{Wall-Collision Effect on Optically-Polarized Atoms in Small and Hot
Vapor Cells}
\author{Yue Chang}
\email{yuechang7@gmail.com}
\author{Jie Qin}

\affiliation{Beijing Automation Control Equipment Institute, Beijing 100074,
China}
\affiliation{Quantum Technology R$\&$D Center of China Aerospace
Science and Industry Corporation, Beijing 100074, China}

\begin{abstract}
In atomic vapor cells, atoms collide with the inner surface, causing their
spin to randomize on the walls. This wall-depolarizing effect is diffusive,
and it becomes more pronounced in smaller vapor cells under high
temperatures. In this work, we investigate the polarization of
optically-pumped alkali-metal atoms in a millimeter-sized cell heated to $%
150 $ Celsius. We consider two extreme boundary conditions: fully
depolarizing and nondepolarizing boundaries, and we provide an analytical
estimation of the polarization difference between them. In the
nondepolarizing case, the pump beam's absorption is proportional to the
average atomic polarization. However, for fully depolarizing walls, the
absorption peak may correspond to a polarization minimum. To mitigate the
wall effect, we propose reducing the pump beam's diameter while maintaining
the pump power to prevent illumination of the cell wall and increase the
pump intensity in the central area. This is crucial for compact vapor-cell
devices where the laser frequency can not be detuned since it is locked to
the absorption peaks. Additionally, we analyze the wall-depolarizing effect
on the performance of an alkali-metal atomic magnetometer operating in the
spin-exchange relaxation-free regime. We show that the signal strength is
highly limited by wall collisions, and we provide an upper bound for it.
\end{abstract}

\maketitle

%\begin{CJK*}{GBK}{song}

%\end{CJK*}

\section{Introduction}

\label{s1}

Optically polarized atoms play a crucial role in various fields, including
magnetometry \cite{Kominis2003,Budker2003,Budker2007}, noble gas
hyper-polarization \cite%
{PhysRevA.29.3092,Walker1997,PhysRevA.58.1412,PhysRevLett.93.160801}, atomic
clocks \cite{vanier2015quantum,forman1990quantum}, quantum optics, and
quantum information \cite%
{julsgaard2004experimental,hammerer2010quantum,moller2017quantum}, as well
as fundamental physics \cite%
{fortson2003search,PhysRevA.75.063416,roberts2015}. Typically, alkali-metal
atoms are confined in a vapor cell and illuminated by a laser beam to
transfer the light's polarization to the atoms' spins \cite%
{happer2010optically,auzinsh2010optically}. However, the atomic polarization
can be limited by collisions with the inner cell walls. In particular, the
spin of alkali-metal atoms can become almost fully randomized when colliding
with the bare glass wall of the cell \cite{RevModPhys.93.035006}, and this
depolarization effect can spread to other parts of the vapor cell through
atom-atom collisions. This boundary effect on the overall polarization is
more significant in small and hot cells. To mitigate this depolarization
effect, antirelaxation coatings with a chemically inert substance such as
paraffin can be used. However, paraffin is not stable even at room
temperature, and thus, it is not widely used in commercial applications. The
development of coatings that can work effectively and remain stable at high
temperatures is still a challenge \cite%
{seltzer2008developments,seltzer2009high,RevModPhys.93.035006}.

In uncoated vapor cells, ways to mitigate the wall-depolarizing effect
include the use of buffer gas, such as nitrogen gas to slow down the motion
of atoms by increasing the time spent colliding with the buffer gas, and
thus the rate at which they reach the cell wall is reduced \cite%
{happer2010optically}. Detuning the pump laser can also reduce the wall
effect by decreasing the optical depth and thus the light absorbed by the
walls \cite{PhysRevA.49.3854}. Most studies on how wall collision affects
polarization focus on large (centimeter scale) or low-temperature cells,
where the wall effect is treated as a constant contribution $\Gamma _{wall}$
to the longitudinal decay rate $\Gamma _{1}$, accounting for the slowest
diffusion mode \cite{knize1988optical,seltzer2008developments}. In this
paper, we systematically study the wall-collision effect on small
(millimeter scale) and hot ($150$ Celsius) alkali-metal vapor cells. By
numerically solving the diffusion equation \cite%
{PhysRev.115.850,PhysRevA.58.1412} with the depolarizing boundary condition,
we find that the wall effect may be underestimated when considering only the
slowing diffusion mode. We compare the average polarization $P_{ave}$ in
depolarizing-wall cells with that in nondepolarizing-wall cells, obtaining
an analytical ratio between these two polarizations as a function of the
diffusion constant $D$, the longitudinal decay rate $\Gamma _{1}$, the
optical pumping rate $R_{op}$, and the system size $L$. The polarization for
depolarizing walls can be much smaller than that with nondepolarizing walls
even if $\Gamma _{wall}$ is added to $\Gamma _{1}$ in the
nondepolarizing-wall case.

The laser beam's propagation is also considered. We prove that the
polarization $P_{ave}$ in nondepolarizing-wall cells illuminated by
uniformly distributed laser beams is independent of the diffusion constant,
and it is proportional to the pump laser's absorption. Thus, $P_{ave}$ can
be acquired from the absorption spectrum. However, this relation does not
hold for depolarizing boundaries, where an absorption peak may correspond to
a local minimum of polarization due to light absorption by the depolarizing
wall. We define a quantity, $\eta _{loss}$, to characterize the pump beam's
loss on the cell walls, which decreases when the pump laser is detuned from
the corresponding transition \cite{PhysRevA.49.3854}. Reducing the diameter
of the laser beam with the same input power also weakens the
wall-depolarizing effect on the overall cell, resulting in more polarization
achieved with a relatively small input power. This is advantageous to
compact vapor-cell based devices where the pump laser's frequency is locked
to the absorption peaks.

Finally, we study the wall-depolarization effect on spin-exchange
relaxation-free (SERF) magnetometers \cite%
{Happer1977,PhysRevLett.89.130801,PhysRevA.71.023405}, which are one of the
most sensitive magnetic-field sensors that typically operate at high
temperatures. When integrating an array of these sensors, the spatial
distribution of the magnetic field can be detected, requiring a small volume
to enhance spatial resolution. By exposing the SERF magnetometer to a small
transverse magnetic field, the linear response $P_{x}$ is extracted. If the
atoms are homogeneously polarized, the transverse signal $P_{x}$ is maximum
when the longitudinal polarization $P_{ave}$ is 1/2 \cite%
{seltzer2008developments}. However, including diffusion, the largest $P_{x}$
occurs at a smaller $P_{ave}<1/2$ for depolarizing boundaries. Moreover,
when the wall is perfectly coated to prevent depolarization, the signal $%
P_{x}$ can increase by an order of magnitude, even with the
depolarizing-wall cell filled with thousands of Torr of nitrogen gas. We
provide an upper bound of the optimal $P_{x}$ for the depolarizing-wall
case, which facilitates efficient signal magnitude estimation without having
to solve diffusion equations for atomic spins and Maxwell equations for
light propagation. We also find that reducing the illuminated area can
enhance the transverse signal $P_{x}$.

The organization of this paper is as follows. In Section \ref{s2}, we begin
by examining the case where the vapor cell is uniformly illuminated. Using $%
\left. ^{87}\text{Rb}\right. $ as an example, we solve the atomic diffusion
equation and the light-propagation equation under two extreme boundary
conditions: fully depolarizing and nondepolarizing walls. We present the
distribution of the polarization and the light intensity in a cylindrical
cell, as well as an analysis of the average polarization and the light
transmission. Additionally, we define the portion $\eta _{loss}$ of light
absorbed by the cell walls. In Section \ref{s3}, we investigate the
transverse signal $P_{x}$ of a SERF magnetometer under various physical
conditions and analyze the optimal $P_{x}$. Moving on to a partially
illuminated vapor cell, Section \ref{s4} examines the dependence of the
polarization $P_{ave}$, the light transmission, and the performance of SERF
magnetometers on the diameter of the laser beam. Finally, in Section \ref{s5}%
, we summarize our work.

\section{Atomic polarization and light transmission}

\label{s2}

In this section, we will solve the light propagation equation along with the
atomic diffusion equation and obtain the light transmission and the atom's
polarization. Without loss of generality, we assume that the vapor cell is
cylindrical with a length of $L$ and a radius of $R$, as shown in Fig. \ref%
{fig1}. The pump beam propagates in the $z$ direction and has a
cross-section radius of $r_{L}$. The longitudinal axis of the cylindrical
cell is the same as the laser beam's, so the system has cylindrical
symmetry. Therefore, the polarization and light intensity $I$ within the
cell are functions of the longitudinal coordinate $z\in \left[ 0,L\right] $
and radial distance $r\in \left[ 0,R\right] $. In the high-temperature
limit, the atomic state can be well approximated by the spin-temperature
distribution \cite{Happer1977,PhysRevA.58.1412} and we can characterize the
atomic polarization $\left\langle F_{z}\right\rangle $ by the electronic
polarization $P_{e}\equiv 2\left\langle S_{z}\right\rangle $ and the
slow-down factor $q\left( P_{e}\right) \equiv \left\langle
F_{z}\right\rangle /\left\langle S_{z}\right\rangle $ is a function of $P_{e}
$. Here, $F_{z}$ ($S_{z}$) is the total (electronic) spin in the $z$
direction. After eliminating adiabatically the excited states \cite%
{PhysRevA.99.063411}, $P_{e}\left( z,r\right) $ in the steady state
satisfies the following equation
\begin{equation}
D\nabla ^{2}q\left( P_{e}\right) P_{e}-\left( R_{op}+\Gamma _{rel}\right)
P_{e}+R_{op}=0,  \label{1}
\end{equation}%
where $D$ is the diffusion constant, $\Gamma _{rel}$ is the atom-collision
induced spin-destruction rate, and $R_{op}=g_{P}\mathcal{L}\left( \Delta
\right) I$ is the optical pumping rate. Here, $g_{P}$ is the square of the
dipole moment $\left\langle d\right\rangle $ and $\mathcal{L}\left( \Delta
\right) $ the lineshape, which will be explicitly given later. Note that we
assume the collisions are strong enough so that the atoms are in a local
spin-temperature distribution with the spin temperature varying as a
function of the atom's coordinate $\left( z,r\right) $. Additionally, we
fill the cell with a buffer gas, such as nitrogen gas to weaken the wall
effect and suppress radiation trapping \cite{molisch1998radiation}. This
ensures that photons emitted from atomic excited states are mostly absorbed
by $N_{2}$, making it possible to neglected the reflection of the laser
beam. Consequently, we can linearize the forward propagation equation of
light as follows:%
\begin{equation}
\partial _{z}I=-g_{I}\mathcal{L}\left( \Delta \right) I\left( 1-P_{e}\right)
.  \label{2}
\end{equation}%
Here, $g_{I}=n_{A}k\left\langle d\right\rangle ^{2}$, where $n_{A}$ is the
density of alkali atoms and $k$ is the laser's wavelength.
\begin{figure}[tbp]
\begin{center}
\includegraphics[width=0.85\linewidth]{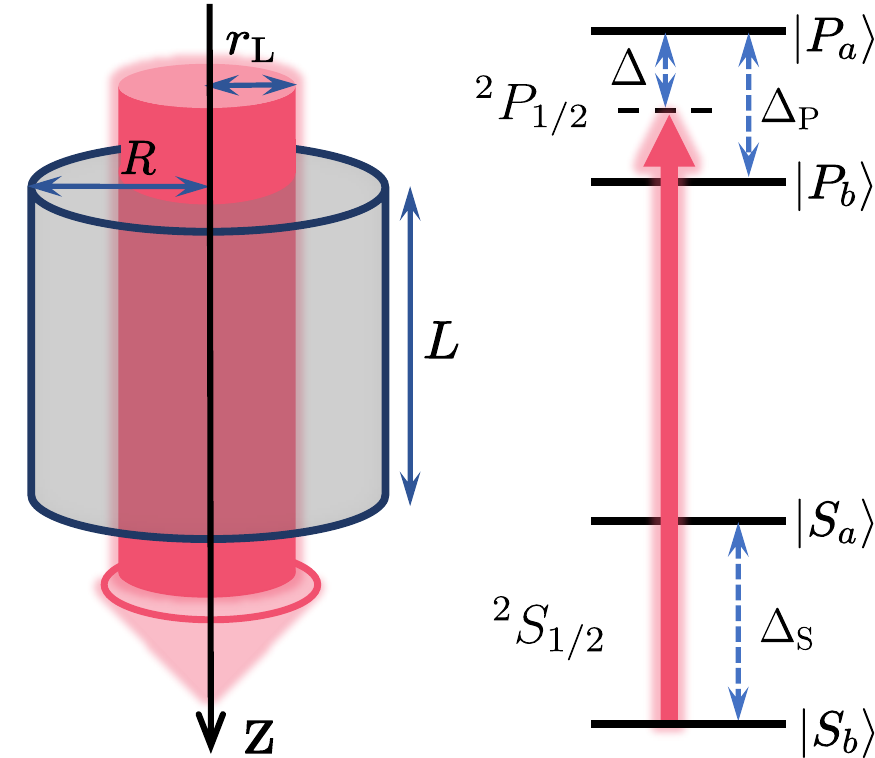}
\end{center}
\caption{Schematic of a cylindrical vapor cell (grey cylinder) optically
pumped by a laser beam (red beam) with a radius of $r_{L}$. The cell has a
longitudinal length of $L$ and a radius of $R$ and is filled with
alkali-metal atoms and buffer gas. The energy level of the alkali-metal atom
is shown on the right. The pump laser couples to the ground states $\left.
^{2}S_{1/2}\right. $ and the first excited states $\left. ^{2}P_{1/2}\right.
$ (D1 transition) with a detuning of $\Delta $. Due to the hyperfine
interaction, the ground (excited) states are split into two manifolds $%
\left\vert S_{a}\right\rangle $ and $\left\vert S_{b}\right\rangle $ ($%
\left\vert P_{a}\right\rangle $ and $\left\vert P_{b}\right\rangle $),
respectively, with an energy difference of $\Delta _{S}$ ($\Delta _{P}$).}
\label{fig1}
\end{figure}

\begin{figure}[tbp]
\begin{center}
\includegraphics[width=0.85\linewidth]{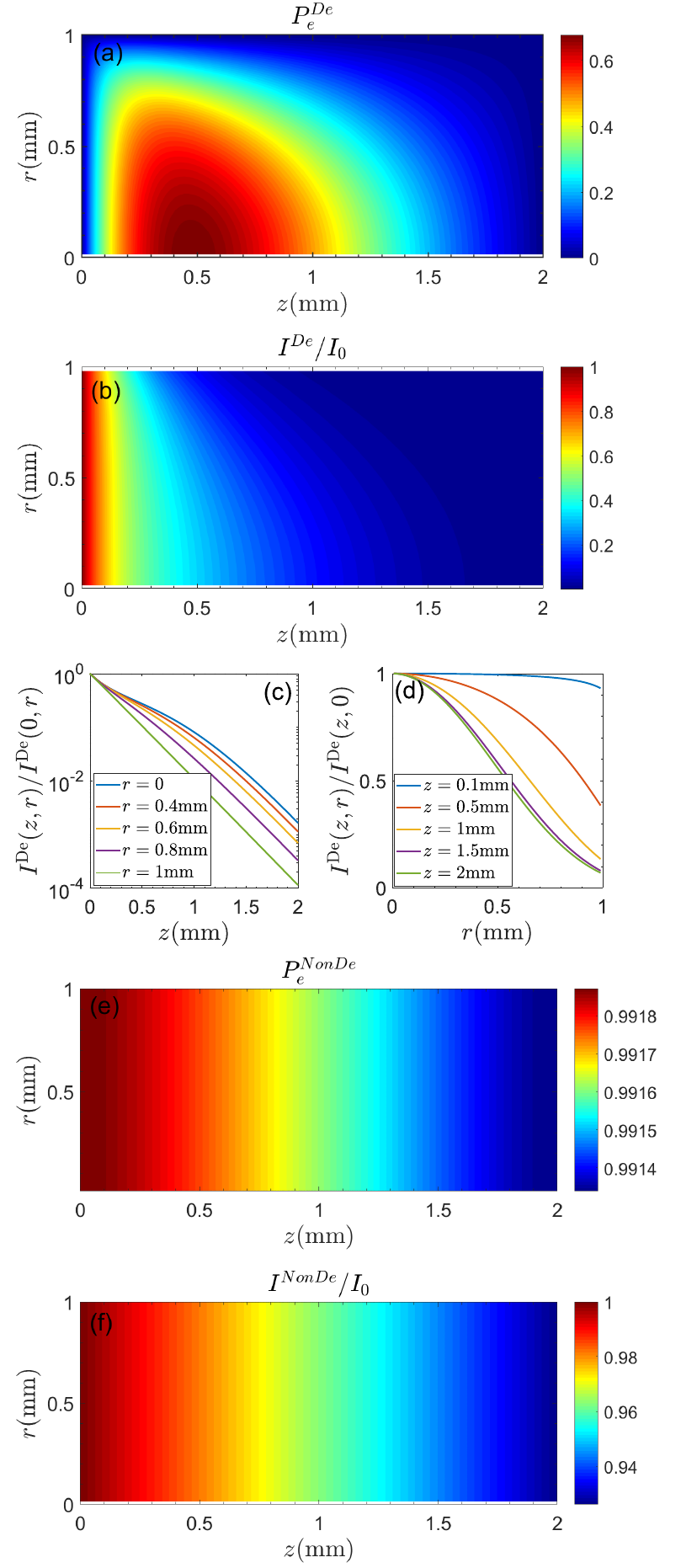}
\end{center}
\caption{Polarization $P_{e}\left( z,r\right) $ and light intensity $I\left(
z,r\right) $ distributions within the vapor cell for depolarizing (a)-(d)
and nondepolarizing (e)-(f) boundaries.}
\label{fig2}
\end{figure}

For $\left. ^{87}\text{Rb}\right. $, the slow-down factor%
\begin{equation}
q\left( P_{e}\right) =\frac{6+2P_{e}^{2}}{1+P_{e}^{2}}
\end{equation}%
and the lineshape can be described as a sum of four Lorentzian functions
corresponding to four transitions between two ground-state and two
excited-state manifolds. This is given by
\begin{eqnarray}
\mathcal{L}\left( \Delta \right)  &=&\frac{\Gamma _{L}}{16\pi }[\frac{5}{%
\Delta ^{2}+\Gamma _{L}^{2}}+\frac{5}{\left( \Delta -\Delta _{S}-\Delta
_{P}\right) ^{2}+\Gamma _{L}^{2}}  \notag \\
&&+\frac{5}{\left( \Delta -\Delta _{S}\right) ^{2}+\Gamma _{L}^{2}}+\frac{1}{%
\left( \Delta -\Delta _{P}\right) ^{2}+\Gamma _{L}^{2}}],
\end{eqnarray}%
where $\Gamma _{L}$ is the decay rate of the excited states and is
proportional to the density of nitrogen gas. We consider the conditions: $%
150^{\circ }$C, $200$Torr $N_{2}$, $L=2R=2$mm, and $\Delta =0$, and plot the
polarization $P_{e}\left( z,r\right) ${}and normalized light intensity $%
I\left( z,r\right) /I_{0}$ in Fig. \ref{fig2}. Here, $I_{0}\equiv I\left(
0,0\right) $ represents the laser intensity at the incident plane, and the
input laser power is $0.5$mW, illuminating uniformly on the entire cell ($%
r_{L}=R$). For depolarizing walls, the polarization $P_{e}^{De}\left(
z,r\right) $ (Fig. \ref{fig2}(a)) is maximum at the center $r=0$ and decays
towards the boundaries due to depolarization of the walls and decrease in
light density. The decrease in light intensity is shown in Fig. \ref{fig2}%
(b), where the intensity $I^{De}\left( z,r\right) $ monotonically decreases
along the propagation direction and decays from the center to the
boundaries. To illustrate this explicitly, Figs. \ref{fig2}(c) and \ref{fig2}%
(d) show the decay of $I^{De}\left( z,r\right) $ along the radical and
longitudinal directions, respectively. For comparison, we show $%
P_{e}^{NonDe}\left( z,r\right) $ and $I^{NonDe}\left( z,r\right) /I_{0}$ for
nondepolarizing walls in Figs. \ref{fig2}(e) and \ref{fig2}(f),
respectively. Here, $P_{e}^{NonDe}\left( z,r\right) $and $I^{NonDe}\left(
z,r\right) $ change only along the pump laser's propagation direction, and
the atomic spins are nearly fully polarized (\symbol{126}$0.99$), leading to
a close-to-unity $I^{NonDe}\left( z,r\right) /I_{0}$.

By integrating Eqs. (\ref{1}) and (\ref{2}) we obtain
\begin{equation}
1-T=\frac{g_{I}}{I_{in}g_{P}}\left[ V\Gamma _{rel}P_{ave}-D\oint d\mathbf{%
S\cdot }\nabla \left( q\left( P_{e}\right) P_{e}\right) \right] ,  \label{8}
\end{equation}%
where%
\begin{equation}
T=\frac{\int_{z=L}\text{d}S\text{ }I\left( r,L\right) }{I_{in}},
\end{equation}%
represents the transmission probability, $S$ is the cell wall,
\begin{equation}
I_{in}=\int_{z=0}\text{d}S\text{ }I\left( r,0\right) ,
\end{equation}%
depicts the total input power, while $P_{ave}$ is the average polarization
over the cell volume $V$:
\begin{equation}
P_{ave}=\frac{1}{V}\int_{V}dV\text{ }P_{e}.
\end{equation}%
If the wall does not depolarize spins ($\left. \nabla P_{e}\right\vert _{S}=0
$), we obtain a simplified expression:
\begin{equation}
1-T=\frac{V\Gamma _{rel}g_{I}}{I_{in}g_{P}}P_{ave}  \label{3}
\end{equation}%
since $\nabla \left( q\left( P_{e}\right) P_{e}\right) =\partial
_{P_{e}}\left( q\left( P_{e}\right) P_{e}\right) \nabla P_{e}$. This implies
that the absorbed light is entirely transferred to the spin polarization $%
P_{ave}$ and that they have a one-to-one correspondence. Additionally, it
can be proved (see Appendix \ref{A1}) that the absorption/polarization peak
occurs at $\partial _{\Delta }\mathcal{L}\left( \Delta \right) =0$.

\begin{figure}[tbp]
\begin{center}
\includegraphics[width=0.85\linewidth]{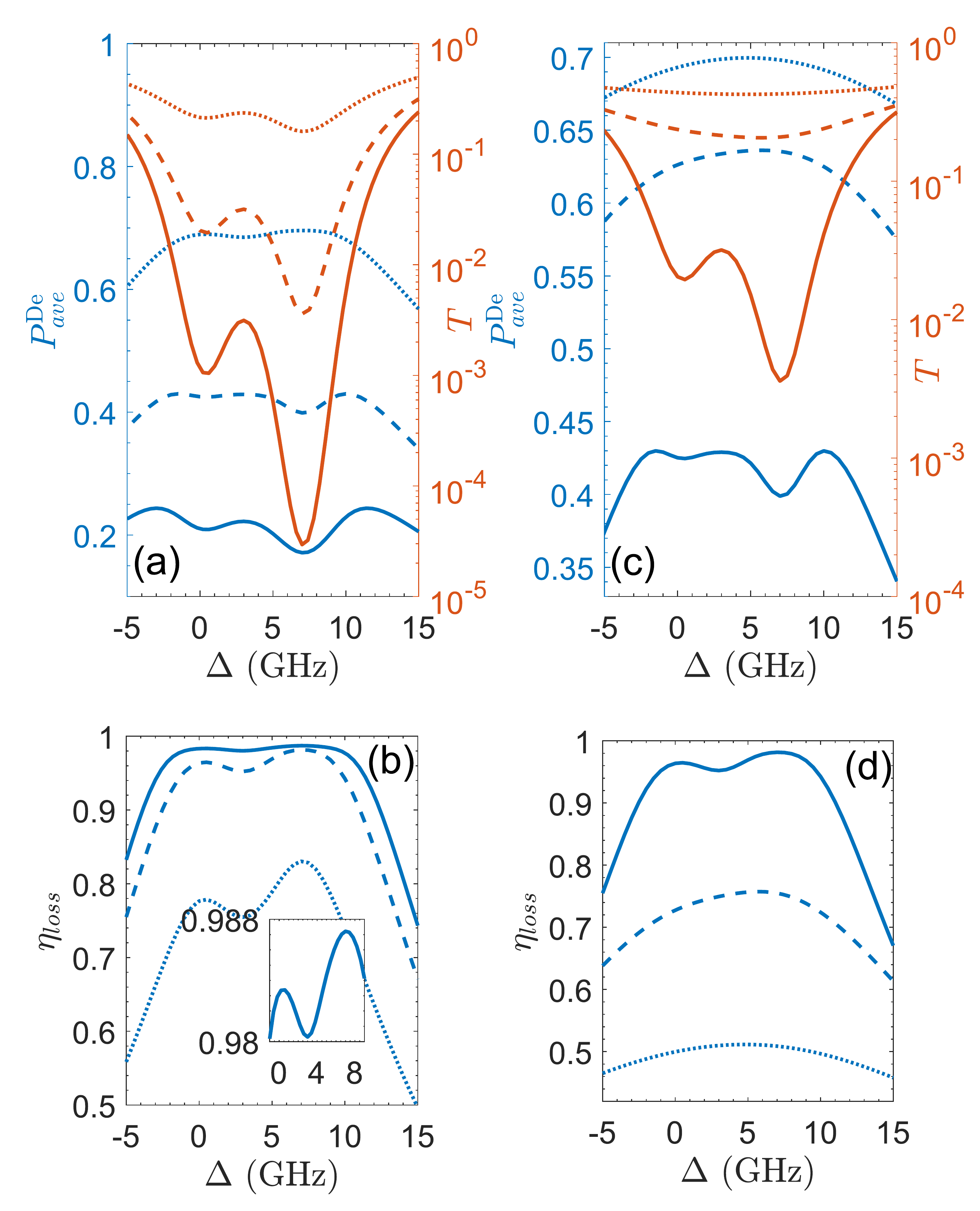}
\end{center}
\caption{Average polarization $P_{ave}$ (blue lines in (a) and (c)),
transmission probability $T$ (red lines in (a) and (c)), and the ratio $%
\protect\eta _{loss}$, representing the fraction of pump laser absorbed by
the cell walls ((b) and (d)) for depolarizing walls. In (a) and (b), the $%
N_{2}$ pressure is $200$Torr, and the input powers are $0.5$mW (solid
lines), $1$mW (dash lines), and $3$mW (dotted lines). In contrast, in (c)
and (d), the input power is fixed at $1$mW, while the $N_{2}$ pressures are
varied from $200$Torr (solid lines), to $500$Torr (dash lines) and $1000$%
Torr (dotted lines). The inset in (b) explicitly shows the two absorption
peaks of the light by the depolarizing walls. All other parameters are
identical to those in Fig. 2.}
\label{fig3}
\end{figure}

In the case of depolarizing walls, $\left. P_{e}\right\vert _{S}=0$ and the
relation (\ref{3}) no longer holds. At the cell walls, $\nabla \left(
q\left( P_{e}\right) P_{e}\right) $ is negative in the direction of $d%
\mathbf{S}$ ($\partial _{P_{e}}\left( q\left( P_{e}\right) P_{e}\right)
=2\partial _{P_{e}}\left\langle F_{z}\right\rangle >0$), and $\oint d\mathbf{%
S\cdot }\nabla \left( q\left( P_{e}\right) P_{e}\right) <0$. Consequently,
more light is absorbed for depolarizing boundaries, but some of it does not
transfer to the spin polarization and is lost in the cell walls. To quantify
this loss, we define the portion of light lost $\eta _{loss}$ as follows:
\begin{equation}
\eta _{loss}=1-T-\frac{V\Gamma _{rel}g_{I}}{I_{in}g_{P}}P_{ave}.
\end{equation}%
For different $N_{2}$ pressures and input powers, we show the average
polarization $P_{ave}^{De}$, the transmission probability $T$, and the loss $%
\eta _{loss}$ in Fig. \ref{fig3}. For small gas pressures ($200$Torr) and
input powers ($<3$mW), the transmission $T$ has two minima at $\Delta
\approx 0$ and $\Delta \approx \Delta _{S}$, respectively (the energy
splitting in the excited states can not be distinguished because $\Gamma
_{L}\sim \Delta _{P}$), while the polarization $P_{ave}^{De}$ is minimal at
these two absorption peaks (Figs. \ref{fig3}(a)). This is due to the large
loss of light on the walls, as shown in Figs. \ref{fig3}(b), where the
absorption of light by the wall is maximal at the resonant points $0$ and $%
\Delta _{S}$ since the bare absorption length $\lambda _{L}\equiv \left[
g_{I}\mathcal{L}\left( \Delta \right) \right] ^{-1}$ of the laser is
shortest (assuming the polarization is zero). This is also true for larger
input powers. When the laser power increases to $3$mW, the wall-depolarizing
effect is reduced since $P_{ave}^{De}$ is larger, so the
absorption-polarization relation is closer to the nondepolarizing-wall case
and the absorption peaks correspond to polarization peaks. Furthermore,
decreasing the diffusion constant $D$ by increasing the $N_{2}$ pressure can
also reduce the wall-depolarizing effect. In Figs. \ref{fig3}(c) and (d),
when the nitrogen gas pressure is increased from $200$Torr\ to $500$Torr and
$1000$Torr, the two absorption peaks, as well as the two peaks in the loss $%
\eta _{loss}$ merge into one peak since $\Gamma _{L}\sim \Delta _{S}$, where
the polarization is maximal. With larger $N_{2}$ pressures, $P_{ave}^{De}$
and $T$ become larger, because the wall-depolarizing effect is smaller (Fig. %
\ref{fig3}(d)) and meanwhile, the absorption length $\lambda _{L}$ is
larger.

For comparison, the polarization and transmission for nondepolarizing walls
are illustrated in Fig. \ref{fig4}. As proved in Appendix A, there exist
absorption/polarization peaks at $\partial _{\Delta }\mathcal{L}\left(
\Delta \right) =0$. For low gas pressure ($200$Torr), the polarization $%
P_{ave}^{NonDe}$ exhibits two peaks at $\Delta \approx 0$ and $\Delta
\approx \Delta _{S}$, corresponding to two transmission minima (Fig. \ref%
{fig4}(a)). With increasing $N_{2}$ pressure, the two peaks become
indistinguishable, and both and the polarization $P_{ave}^{NonDe}$ and the
transmission $T$ decrease (Fig. \ref{fig4}(c)). The decrease in transmission
is due to the increase in $\Gamma _{rel}$, which exceeds the effect of $%
\lambda _{L}$'s increment on the polarization. In the absence of
depolarization on the walls, the transmission $T$ is close to unity and the
polarization $P_{ave}^{NonDe}$ is much larger than $P_{ave}^{De}$.
Therefore, the ratio $P_{ave}^{De}/P_{ave}^{NonDe}$ can also be used to
characterize the wall-depolarization effect, as plotted in Figs. \ref{fig4}%
(b) and (d) in blue lines. Like the light lost $\eta _{loss}$, the ratio $%
P_{ave}^{De}/P_{ave}^{NonDe}$ can be increased by detuneing the pump laser,
increaseing the pump power, or increasing $N_{2}$ density. Apart from
solving the nonlinear equations (\ref{1}) and (\ref{2}), this ratio $%
P_{ave}^{De}/P_{ave}^{NonDe}$ can also be estimated as (see Appendix \ref{A2}
for the derivation)%
\begin{equation}
\frac{P_{ave}^{De}}{P_{ave}^{NonDe}}\approx \sqrt{1-\frac{\lambda _{D}}{%
\lambda _{L}}}\left( 1-2\frac{\lambda _{D}}{L}\right) \left( 1-\frac{\lambda
_{D}}{R}\right) ^{2}  \label{4}
\end{equation}%
where%
\begin{equation}
\lambda _{D}=\sqrt{\frac{qD}{\Gamma _{rel}+R_{op}^{0}}.}
\end{equation}%
Here, $\lambda _{D}$ is the wall-depolarization length with $%
R_{op}^{0}\equiv g_{P}\mathcal{L}\left( \Delta \right) I_{0}$. The validity
of this estimation requires the ratio $\lambda _{D}/\lambda _{L}$ between
the wall-depolarization length $\lambda _{D}$ and the absorption length $%
\lambda _{L}$ is smaller than $1$. The smaller $\lambda _{D}/\lambda _{L}$,
the better the analytical estimation, as shown in Figs. \ref{fig4}(b) and
(d) in red lines. The ratio $\lambda _{D}/\lambda _{L}$ can be reduced by
increasing the incident light density $I_{0}$ or the detuning $\Delta $ from
the resonant points, or decreasing the diffusion constant $D$. When $\lambda
_{D}/\lambda _{L}$ is not sufficiently small so that the approximation (\ref%
{4}) is not perfect (solid lines in Figs. \ref{fig4}(b) and (d)), but it
still provides an upper bound for $P_{ave}^{De}/P_{ave}^{NonDe}$, since we
have assumed the light density $I$ does not change much within the
wall-depolarization length and $\lambda _{D}$ consequently depends solely on
the initial value $I_{0}$. Note that the slow-down factor $q$ is a function
of the local polarization $P_{e}\left( z,r\right) $ but a median value can
be used to simplify the calculation. For $\left. ^{87}\text{Rb}\right. $, $q$
varies between $4$ (fully polarized) and $6$ (zero polarization), and we
take $q=5$ in Eq. (\ref{4}).

\begin{figure}[tbp]
\begin{center}
\includegraphics[width=0.85\linewidth]{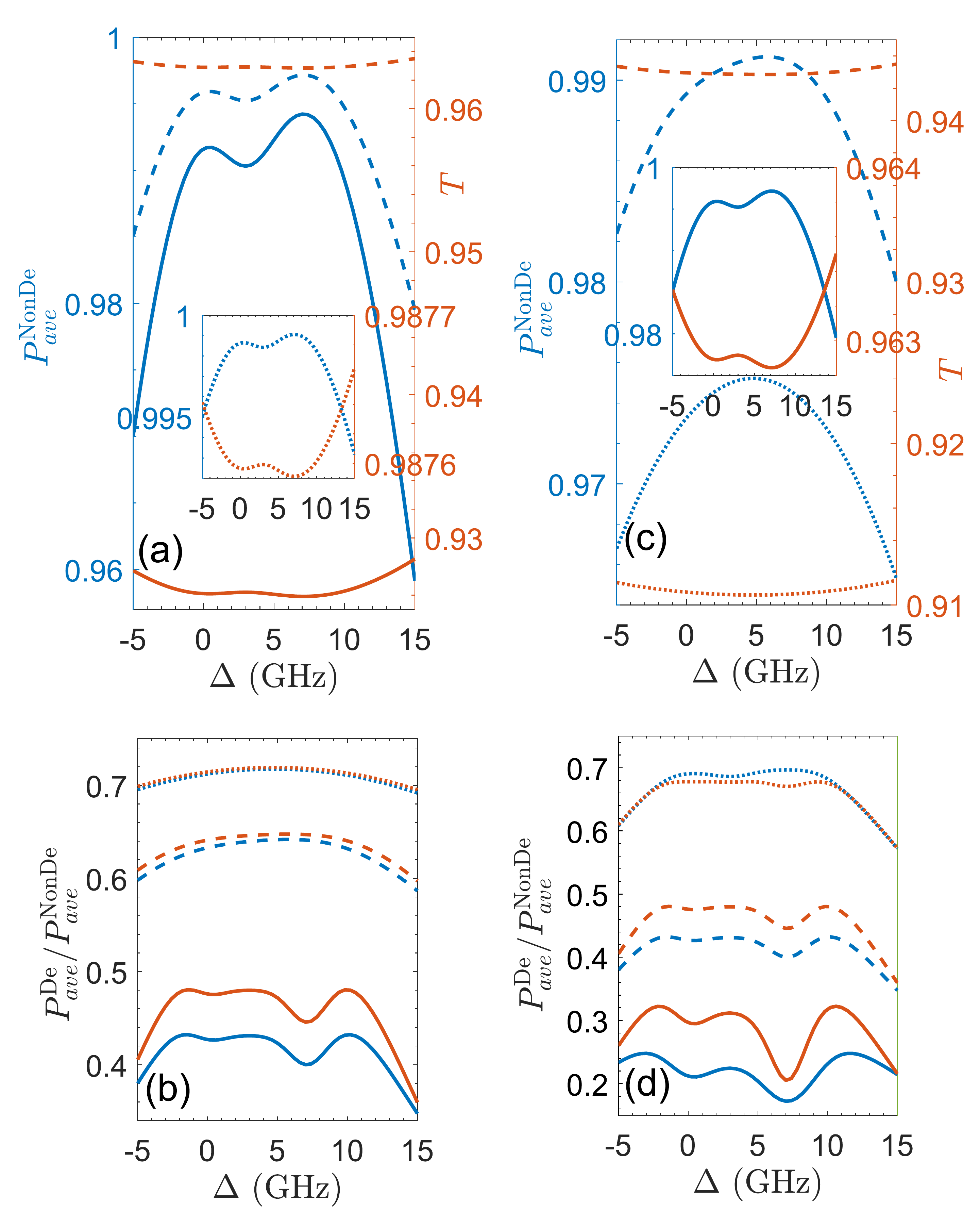}
\end{center}
\caption{Average polarization $P_{ave}^{NonDe}$ (blue lines in (a) and (c))
and transmission probability $T$ (red lines in (a) and (c)) for
nondepolarizing walls. The ratio $P_{ave}^{De}/P_{ave}^{NonDe}$ is shown in
(c) and (d). Similar to Figs. \protect\ref{fig3}, the $N_{2}$ pressure is
fixed at $200$Torr in (a) and (b), while input powers of $0.5$mW (solid
lines), $1$mW (dash lines), and $3$mW (dotted lines) are used. In (c) and
(d), the input power is fixed at $1$mW, and $N_{2}$ pressures of $200$Torr
(solid lines), $500$Torr (dash lines), and $1000$Torr (dotted lines) are
used. The remaining parameters are identical to those in Fig.2. The
numerical results are shown with blue lines in (b) and (d), while the
analytical estimation using Eq. (\protect\ref{4}) is shown with red lines.
In this case, the slow-down factor $q$ in Eq. (\protect\ref{4}) is fixed at $%
5$.}
\label{fig4}
\end{figure}

Literally, a commonly used approach to account for the depolarizing-wall
effect is to introduce a spin-depolarization rate $\Gamma _{wall}$, which is
determined by the lowest diffusion mode, to the longitudinal relaxation
rate. In this study, we investigate this method by substituting $\tilde{%
\Gamma}_{{rel}}=\Gamma _{{rel}}+\Gamma _{wall}$ for the nondepolarizing-wall
case. For cylindrical cells,
\begin{equation}
\Gamma _{wall}=qD\left[ \left( \frac{\pi }{L}\right) ^{2}+\left( \frac{\mu
_{1}}{R}\right) ^{2}\right] ,  \label{5}
\end{equation}%
where $\mu _{1}$ is the first zero of the Bessel function of the first kind
\cite{PhysRevA.49.3854,lee2022ultra}. By taking the maximum value of $q=6$
in Equation (\ref{5}), we depict $\tilde{P}_{{ave}}^{NonDe}$ in Figure \ref%
{fig5} with red lines, while $P_{{ave}}^{De}$ is shown with blue lines for
comparison. Our findings reveal that, even with the inclusion of $\Gamma
_{wall}$ in $\Gamma _{rel}$, the polarization $\tilde{P}_{{ave}}^{NonDe}$
can be significantly larger than $P_{{ave}}^{De}$. Thus, this simplistic
approach may underestimate the wall-depolarizing effect.

\begin{figure}[tbp]
\begin{center}
\includegraphics[width=0.85\linewidth]{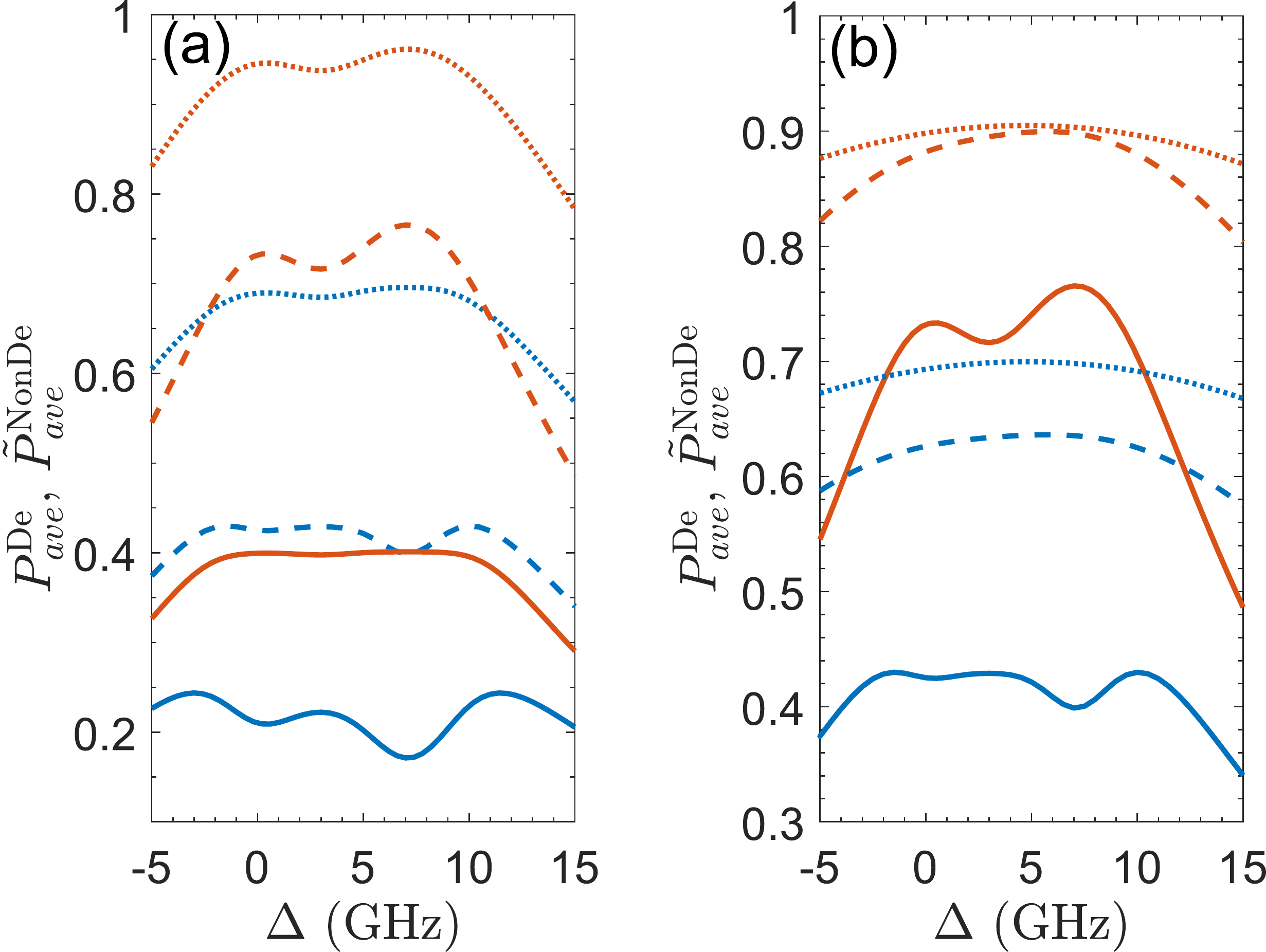}
\end{center}
\caption{Average polarization $\tilde{P}_{ave}^{NonDe}$ (red lines) for
nondepolarizing walls by including $\Gamma _{wall}$ (with slow-down factor $%
q=6$) in the longitudinal decay. In comparison, $P_{ave}^{De}$ is shown in
blue lines. Similar to Figs. \protect\ref{fig3}, (a) displays the results
for $N_2$ pressure of $200$ Torr and input powers of $0.5$ mW (solid lines),
$1$ mW (dashed lines), and $3$ mW (dotted lines). In (b), the input power is
constant at $1$ mW, while the $N_{2}$ pressures vary among $200$Torr (solid
lines), $500$Torr (dash lines), and $1000$Torr (dotted lines). All other
parameters remain unchanged from Fig.2.}
\label{fig5}
\end{figure}

\section{Wall effect on SERF magnetometers}

\label{s3}

In this section, we take SERF magnetometers \cite%
{Happer1977,PhysRevLett.89.130801,Kominis2003} as an example to study the
wall effect on the transverse signals $P_{x}\equiv 2\left\langle
S_{x}\right\rangle $. SERF magnetometers are highly sensitive to magnetic
fields and do not suffer from spin-exchange relaxation, which is a major
source of spin-decoherence in atomic magnetometers. In order to operate in
the SERF regime, SERF magnetometers are typically used at high temperatures
and low magnetic fields $B$. Here, we assume the magnetic field $B$ to be
along the $y$ direction, and its induced energy splitting much smaller than
the spin-destruction rate $\Gamma _{rel}$. As in Sec. \ref{s2}, we assume
that the atomic system is approximately in a spin-temperature distribution.
To the first order of the small magnetic field $B$, $P_{x}$ in the long-term
limit satisfies the following equation%
\begin{equation}
D\nabla ^{2}q\left( P_{e}\right) P_{x}-\left( R_{op}+\Gamma _{rel}\right)
P_{x}+\gamma _{e}BP_{e}=0,  \label{6}
\end{equation}%
where $\gamma _{e}$ is the electron's gyromagnetic ratio and $\gamma
_{e}B\ll \Gamma _{rel}$. Note that here, we have ignored the light shift
induced by the pump laser, as it is usually compensated for by a magnetic
field along the $z$ direction in practical use.

\begin{figure}[tbp]
\begin{center}
\includegraphics[width=0.85\linewidth]{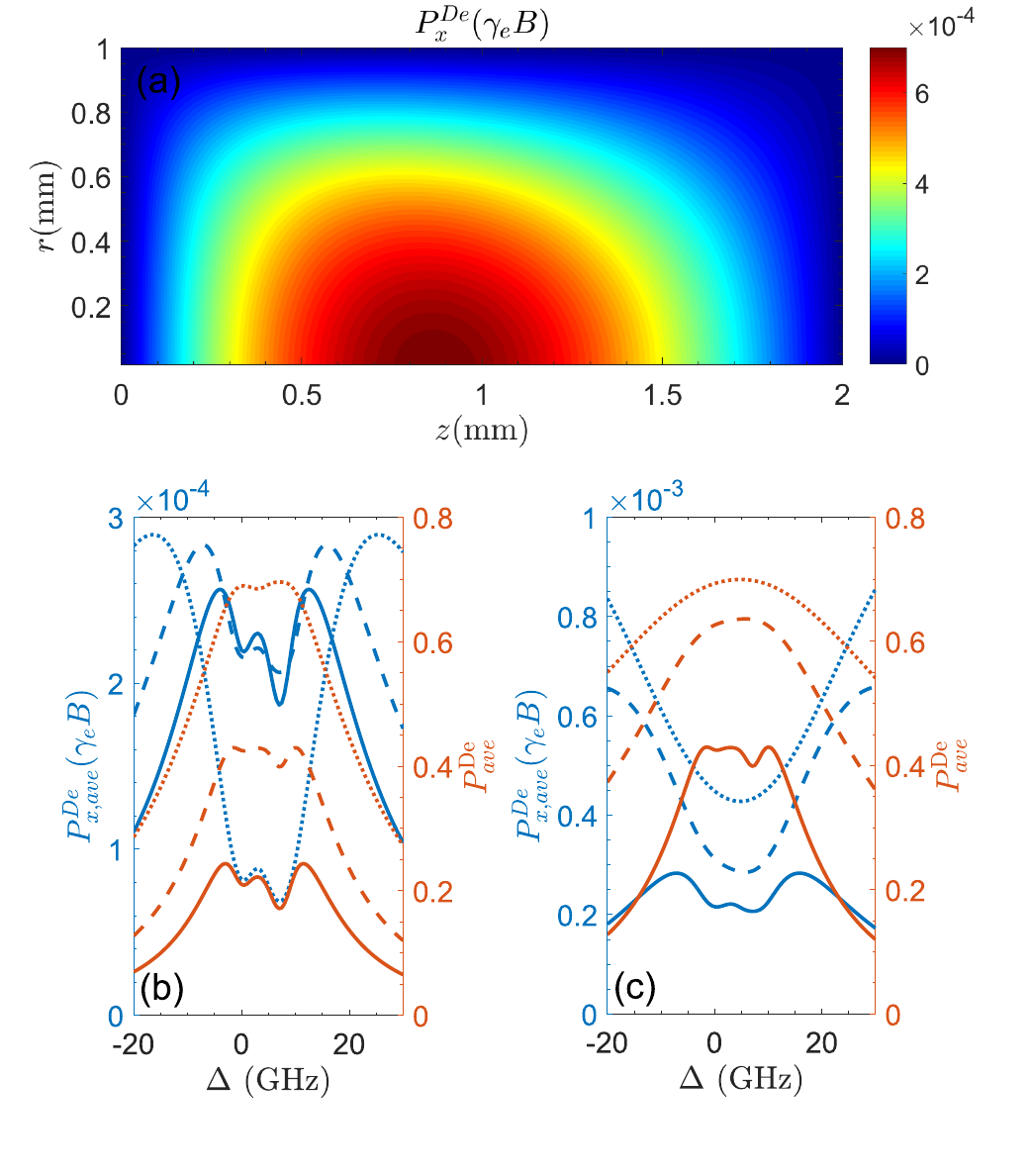}
\end{center}
\caption{The spatial dependence of $P_{x}^{De}\left( z,r\right) $ for
depolarizing walls is displayed in (a), using the same parameters as in
\protect\ref{fig2}(a). In (b) and (c), the average signal $P_{x,ave}^{De}$
is represented by blue lines, while the corresponding polarization $%
P_{ave}^{De}$ is depicted by red lines. Same as in Fig. \protect\ref{fig3},
the $N_{2}$ pressure is $200$Torr and the input powers are $0.5$mW (solid
lines), $1$mW (dashed lines), and $3$mW (dotted lines) in (b). In (c), the
input power is $1$mW, and the $N_{2}$ pressures are $200$Torr (solid lines),
$500$Torr (dashed lines), and $1000$Torr (dotted lines). All other
parameters remain the same as in Fig. 2.}
\label{fig6}
\end{figure}

The solution $P_{x}^{De}\left( z,r\right) $\ to Eq. (\ref{6}) for
depolarizing boundaries is depicted in Fig. \ref{fig6}(a), where the
parameters used are the same as in Fig. \ref{fig2}(a). It can be observed
that Fig. \ref{fig6}(a), $P_{x}^{De}\left( z,r\right) $ reaches its maximal
value in the region where $P_{e}^{De}\left( z,r\right) \approx 1/2$, which
is consistent with the simple model of SERF magnetometers that disregards
the light propagation and atomic diffusion. However, it should be noted that
the maximum of the total transverse signal
\begin{equation}
P_{x,ave}\equiv \frac{1}{V}\int_{V}dV\text{ }P_{x}
\end{equation}%
appears at $P_{ave}^{De}<1/2$ (See Figs.\ref{fig6}(b) and (c)). This is
because the polarizations $P_{e}^{De}\left( z,r\right) $ decays as it
approaches the cell walls. For small $P_{ave}^{De}$ (solid lines in Fig.\ref%
{fig6}(b) and (c)), the transverse signal $P_{x,ave}$ locally reaches its
minimum at the absorption peaks, where $P_{ave}^{De}$ is also locally
minimal. This behavior preserves even for large input region, as depicted by
the dash and dotted lines in \ref{fig6}(b) and (c). This is in contrast to $%
P_{ave}^{De}$, since with large input power, $P_{ave}^{De}$ reaches a local
maximum and exceeds $1/2$ at the absorption peaks.

\begin{figure}[tbp]
\begin{center}
\includegraphics[width=0.9\linewidth]{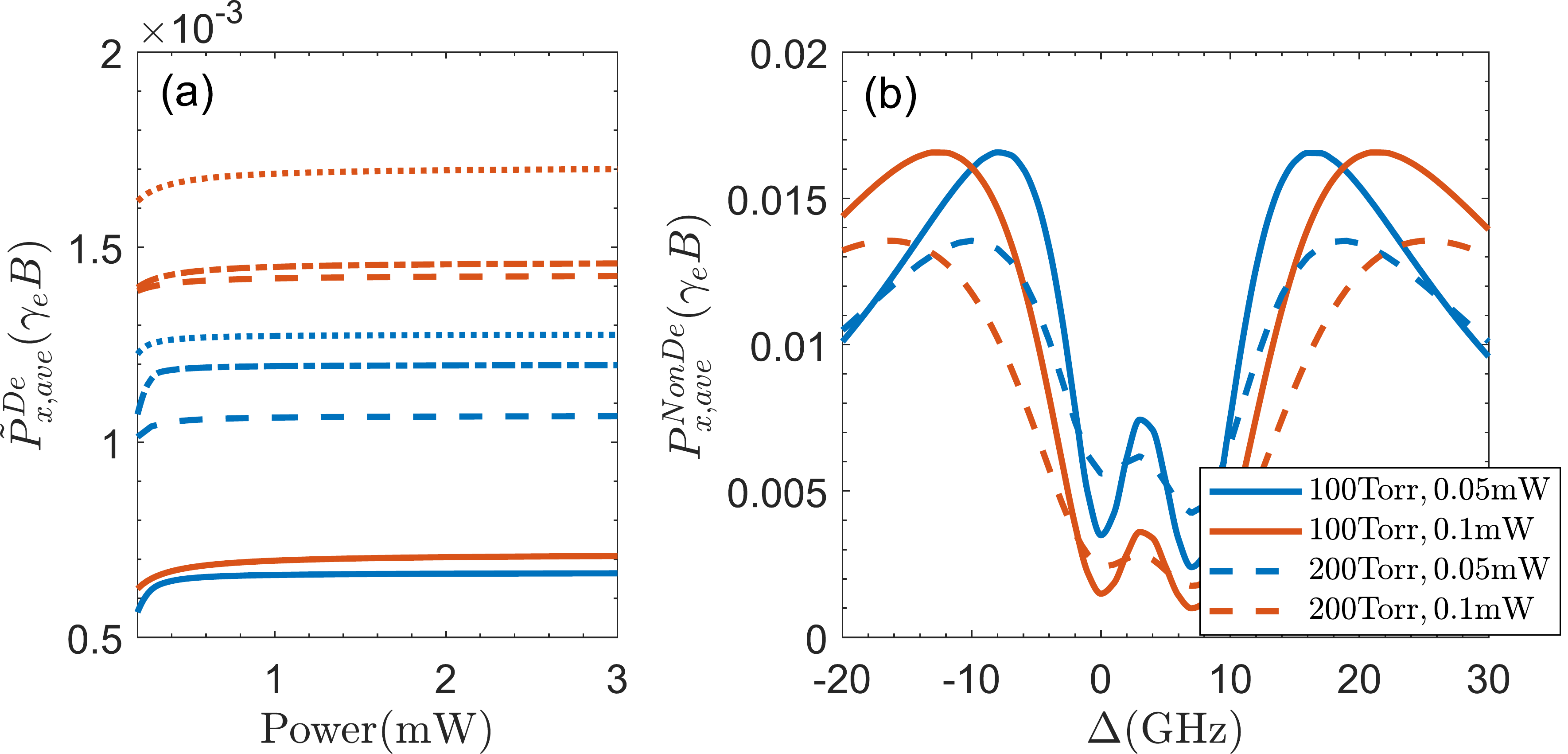}
\end{center}
\caption{(a) Transverse signal $\tilde{P}_{x,ave}^{De}$ as a function of the
pump laser power at various $N_{2}$ pressures. The blue lines represent the
numerical results obtained by varying the detuning $\Delta $, while the red
lines represent the analytical estimation (Eq. (\protect\ref{7})) that
provides an upper bound for $\tilde{P}_{x,ave}^{De}$ with a difference of no
more than $30\%$. The $N_{2}$ pressures considered are $500$ Torr (solid
lines), $1000$ Torr (dashed lines), $2000$ Torr (dotted lines), and $3000$
Torr (dash-dotted lines). The remaining parameters are the same as in Fig.
\protect\ref{fig2}(a). (b) Transverse signal $P_{x,ave}^{NonDe}$ for
nondepolarizing walls at a temperature of $150$ Celsius.}
\label{fig7}
\end{figure}

To increase the total transverse signal $P_{x,ave}$, one can enlarge the
region where the electron polarization $P_{e}^{De}\left( z,r\right) $ is
approximately $1/2$, since the transverse polarization $P_{x}^{De}\left(
z,r\right) $ is large in this region. This can be achieved by increasing the
buffer gas density or by detuning the pump laser while increasing its power
to compensate for the decrease in optical pumping rate. Optimal signals $%
\tilde{P}_{{x,ave}}^{De}$ are obtained for different input powers of the
pump laser by varying the detuning $\Delta $ for different $N_{{2}}$
pressures. As shown in Figure \ref{fig7}(a), for a given $N_{2}$ pressure,
the maximal signal $\tilde{P}_{x,ave}^{De}$ saturates as the input power
increases because the wall-depolarization length $\lambda _{{D}}$ saturates
and the region where $P{e}^{De}\left( z,r\right) \approx 1/2$ does not
extend. With more $N_{2}$, the wall-depolarization length $\lambda _{D}$
saturates at a smaller value and the region where $P_{e}^{De}\left(
z,r\right) \approx 1/2$ is further enlarged, resulting in a larger $\tilde{P}%
_{x,ave}^{De}$ ($\left. \tilde{P}_{x,ave}^{De}\right\vert _{500\text{Torr}%
}<\left. \tilde{P}_{x,ave}^{De}\right\vert _{1000\text{Torr}}<\left. \tilde{P%
}_{x,ave}^{De}\right\vert _{2000\text{Torr}}$). However, the dependence of $%
\tilde{P}_{x,ave}^{De}$ on the $N_{2}$ pressure is not monotonic, because $%
P_{x}^{De}=\gamma _{e}BP_{e}^{De}/\left( R_{op}+\Gamma _{rel}\right) $ when
ignoring the spatial dependence and $\Gamma _{rel}$ is larger with more
buffer gas. Therefore, $\left. \tilde{P}_{x,ave}^{De}\right\vert _{3000\text{%
Torr}}<\left. \tilde{P}_{x,ave}^{De}\right\vert _{2000\text{Torr}}$ since
the increase in $\Gamma _{rel}$ has a greater influence than the suppression
of the wall-depolarization effect.

The optimal transverse signal $\tilde{P}_{x,ave}^{De}$ for depolarizing
walls can be estimated analytically as (see Appendix \ref{A2} for details)%
\begin{equation}
\tilde{P}_{x,ave}^{De}\approx \min \{r_{1},r_{2}\},  \label{7}
\end{equation}%
where%
\begin{equation*}
r_{1}=\max \left[ \frac{P_{ave}^{De}}{P_{ave}^{NonDe}}\frac{\gamma
_{e}BR_{op}^{0}}{\left( R_{op}^{0}+\Gamma _{rel}\right) ^{2}}\right]
\end{equation*}%
and%
\begin{equation}
r_{2}=\max \left[ \frac{P_{ave}^{De}}{P_{ave}^{NonDe}}\frac{\gamma _{e}B}{%
2\left( R_{op}^{0}+\Gamma _{rel}\right) }\right] .
\end{equation}%
This analytical upper bound of $\tilde{P}_{{x,ave}}^{De}$ is shown in Fig. %
\ref{fig7}(a), which has up to $30\%$ deviation from the exact numerical
values. Therefore, using the simple expression in Eq. (\ref{7}) can be
helpful to determine the optimal parameters without solving the nonlinear
equations. It should be noted that even with optimized parameters, the
transverse signal for depolarizing walls can be significantly smaller than
the non-depolarizing wall case. For example, for $100$Torr and $200$Torr $N{2%
}$, which are sufficient to suppress radiation trapping, the latter case is
presented in Fig. \ref{fig7}(b). In this case, the signal $P_{x,ave}^{NonDe}$
can be one order of magnitude larger than $\tilde{P}_{x,ave}^{De}$ due to
the strong wall-depolarizing effect in small vapor cells.

\section{Partially illuminated cells}

\label{s4}

\begin{figure}[tbp]
\begin{center}
\includegraphics[width=0.9\linewidth]{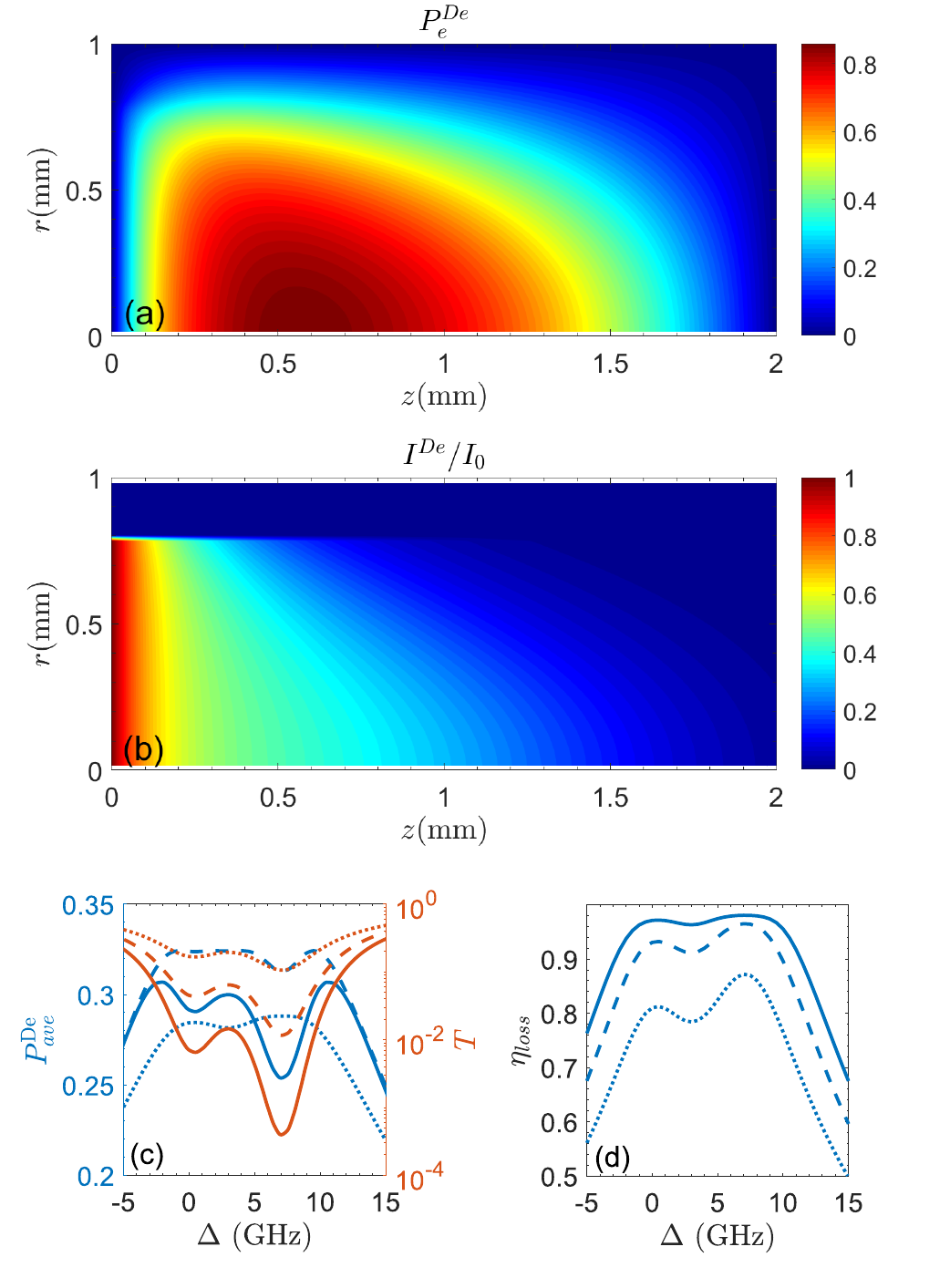}
\end{center}
\caption{Spatially dependent atomic polarization $P_{e}^{De}\left(
z,r\right) $ (a) and normalized light intensity $I^{De}\left( z,r\right)
/I_{0}$ (b) for partially illuminated, uncoated cells with laser beam radius
$r_{L}=0.8$mm, while the other parameters are kept the same as in Fig.
\protect\ref{fig2}(a). The average polarization $P_{ave}^{De}$ (blue lines)
and the transmission probability $T$ (red line) are presented in (c), while
the light loss $\protect\eta _{loss}$ by the depolarizing wall is shown in
(d). Here, the radius $r_{L}$ is $0.8$mm (solid lines), $0.6$mm (dashed
lines), and $0.4$mm (dotted lines), respectively.}
\label{fig8}
\end{figure}

Uncoated cells exhibit depolarization of atoms in the vicinity of the
boundary, within a distance of $\lambda _{D}$. To mitigate this
wall-depolarizing effect, the laser beam can be concentrated on the central
region of the cell while maintaining constant input power or $I_{in}$. To
demonstrate this, we simulate the atomic polarization $P_{e}^{De}\left(
z,r\right) $ and the light intensity $I^{De}\left( z,r\right) /I_{0}$, with
a laser radius of $0.8$mm and input power of $0.5$mW, depicted in Figs. \ref%
{fig8}(a) and (b), respectively. While the light is confined to the central
region of the cell due to radial non-propagation, the atomic polarization in
the un-illuminated region is still non-zero as spins diffuse in all
directions via collisions. Note that the incident light intensity $I_{0}$ is
dependent on the illumination area as $I_{0}=I_{in}/\left( \pi
r_{L}^{2}\right) $. Our results show that the partially illuminated cell
exhibits more centralized atomic polarization and a slower rate of decay of
the light intensity $I^{De}\left( z,r\right) /I_{0}$ along the $z$%
-direction, as compared to Figs. \ref{fig2}(a) and (b). The overall effect
is depicted in Figs. \ref{fig8}(c) and (d), where the average polarization $%
P_{ave}^{De}$ and light transmission $T$ are both increased, resulting in a
reduced light loss $\eta _{loss}$ due to the wall.

The impact of centralizing the laser beam on the wall-depolarizing effect is
most evident at resonant points that correspond to absorption peaks. This is
highly beneficial for compact atomic-vapor-cell based devices, the frequency
of which is locked to absorption peaks. In this study, we examined the
polarization $P_{ave}^{De}$ at the absorption peaks and the transverse
signal $P_{x,ave}^{De}$ of the SERF magnetometer under various input powers
and $N_{2}$ pressures.

\begin{figure}[tbp]
\begin{center}
\includegraphics[width=0.9\linewidth]{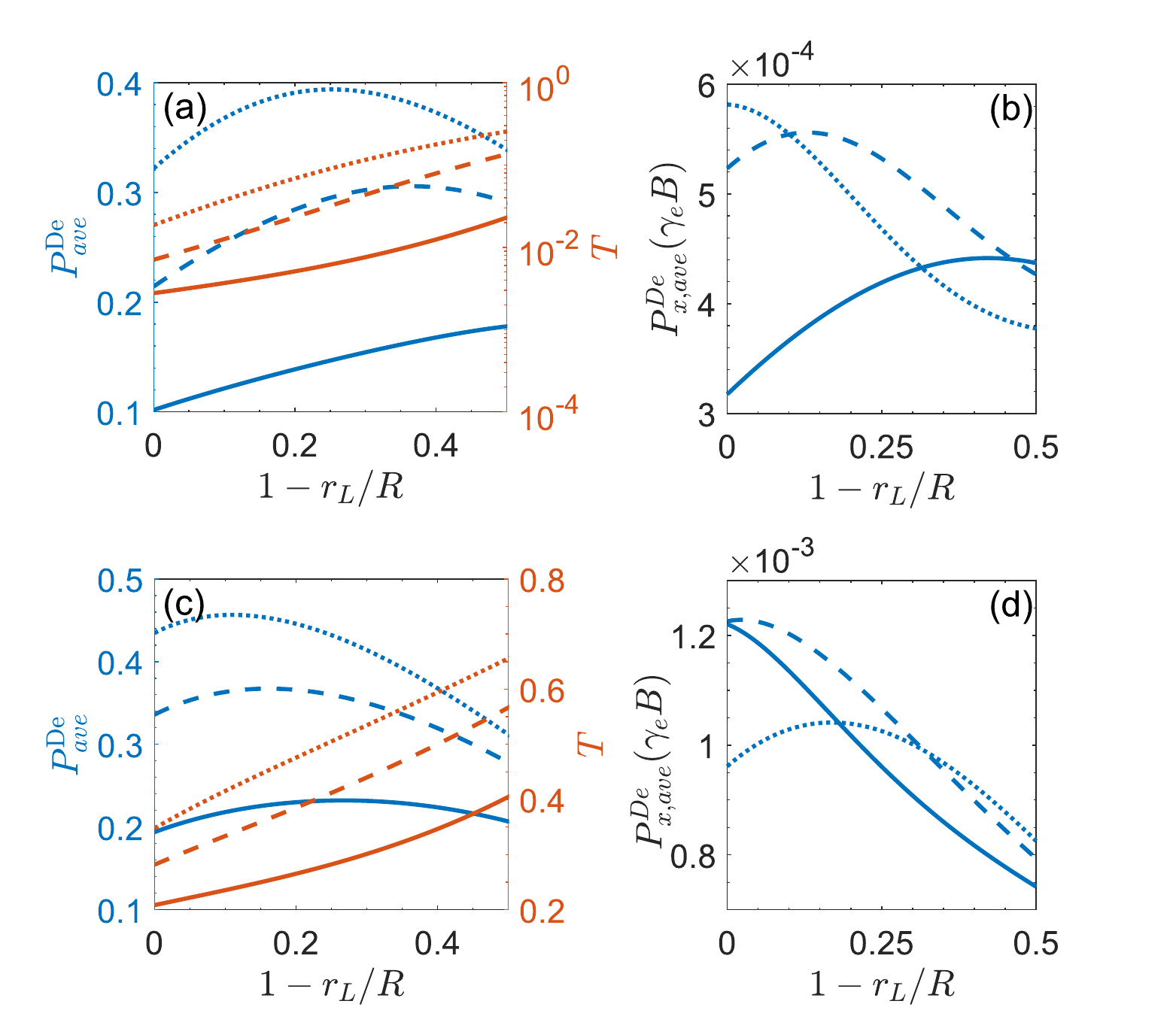}
\end{center}
\caption{Average atomic polarization $P_{ave}^{De}$, pump laser's
transmission probability $T$, and transverse signal $P_{x,ave}^{De}$ of SERF
magnetometers at different $N_{2}$ pressures and input powers. The $N_{2}$
pressures are $500$Torr for (a) and (b) and $2000$Torr for (c) and (d). The
laser input powers are $0.1$mW (solid line), $0.2$mW (dashes line), and $0.3$%
mW (dotted line), respectively.}
\label{fig9}
\end{figure}

As $r_{L}$ decreases, the effective absorption length increases, and the
wall-polarization effect is reduced, leading to a higher polarization $%
P_{ave}^{De}$. However, once the polarization centralized around the
illuminating region approaches its maximum value, the total polarization
begins to decrease. Meanwhile, reducing the light lost due to wall
depolarizing by monotonically decreasing $r_{L}$ results in a significant
increase in the transmission probability. This increase is particularly
noticeable at low $N_{2}$ densities ($500$ Torr in Figs. \ref{fig9}(a) and
(b)). For example, with $0.2$ mW of input power (dashed lines), $P_{ave}^{De}
$ increases from $0.21$ to $0.31$ when the beam radius is reduced from $R$
to $0.65R$, while $T $ increases from $0.007$ to $0.06$. This one-order
increase in transmission probability is advantageous in locking the laser's
frequency. Furthermore, reducing $r_{L}$ also results in an increase in the
signal $P_{x,ave}^{De}$. These characteristics of $P_{ave}^{De}$, $T$, and $%
P_{x,ave}^{De}$ are consistent across different $N_{2}$ densities ($2000$
Torr in Figs. \ref{fig9}(c) and (d)), although their changes with respect to
$r_{L}$ are less significant since the wall-depolarizing effect is smaller
when more buffer gas is present.

\section{Conclusions}

\label{s5} In this work, we have investigated the wall effect in small
atomic vapor cells at high temperatures. By comparing cases with fully
depolarizing and nondepolarizing walls, we found that the traditional
treatment, which considers only the lowest diffusion mode, underestimates
the wall-depolarizing effect in the $2$mm-cells. To address this issue, we
proposed a theoretical estimation of the ratio between the polarizations for
depolarizing and nondepolarizing boundaries.

To demonstrate practical implications of our findings, we focus on the SERF
magnetometer and its transverse signal dependence on the wall effect. We
derived a theoretical upper-bound for the uncoated cells, which revealed
that the optimal signal for cells with depolarizing walls is one order
smaller compared to that of nondepolarizing walls. Our study also presents a
novel approach to reduce the wall-depolarizing effect by shrinking the
beam's radius while keeping its input power, in addition to detuning the
pump laser. This method can enhance various physical quantities in
concerned, such as the atomic polarization, the transmission probability,
and the transverse signal of the SERF magnetometer.

%%%%%%%%%%%%%%%%%%%%%%%%%%%%%%%%%%%%%%%%%%%%%%%%%%%%%%%%%%%%%%%%%%%%%%%
\acknowledgments This work was supported by the National Natural Science
Foundation of China Grants No. U2141237 and No.62101070.
%%%%%%%%%%%%%%%%%%%%%%%%%%%%%%%%%%%%%%%%%%%%%%%%%%%%%%%

\appendix

\section{Absorption/Polarization peaks for nondepolarizing walls}

\label{A1}

In this section, we will prove that the absorption/polarization peaks for
nondepolarizing walls are at $\partial _{\Delta }\mathcal{L}\left( \Delta
\right) =0$.

We first integrate Eq. (\ref{2}) and acquired
\begin{equation}
I\left( L\right) =I\left( 0\right) e^{-g_{I}\mathcal{L}\left( \Delta \right)
\int_{0}^{L}\text{d}z\left[ 1-P_{e}\left( z\right) \right] }.  \label{a1}
\end{equation}%
Note that for nondepolarizing cylindrical boundaries, $I$ and $P_{e}$ are
uniformly distributed along the radial direction, so they are only functions
of $z$. Then the derivative of Eq. (\ref{a1}) $\partial _{\Delta }I\left(
L\right) $ reads
\begin{eqnarray}
&&\partial _{\Delta }I\left( L\right)  \notag \\
&=&g_{I}I\left( L\right) [\mathcal{L}\left( \Delta \right) \partial _{\Delta
}\int_{0}^{L}\text{d}z\text{\thinspace }P_{e}\left( z\right) -  \notag \\
&&\int_{0}^{L}\text{d}z\left[ 1-P_{e}\left( z\right) \right] \partial
_{\Delta }\mathcal{L}\left( \Delta \right) ].
\end{eqnarray}%
At the peaks, $\partial _{\Delta }I\left( L\right) =0$\ and equivalently $%
\partial _{\Delta }\int_{0}^{L}dz$ $P_{e}\left( z\right) =0$ (see Eq. (\ref%
{3}) for uniform irradiation case), which consequently gives%
\begin{equation}
\int_{0}^{L}\text{d}z\left[ 1-P_{e}\left( z\right) \right] \partial _{\Delta
}\mathcal{L}\left( \Delta \right) =0.
\end{equation}%
Since $1-P_{e}\left( z\right) >0$, the maximal polarization/absorption is at
$\partial _{\Delta }\mathcal{L}\left( \Delta \right) =0$.

\section{Estimation of $P_{ave}^{De}/P_{ave}^{NonDe}$ and $\tilde{P}%
_{x,ave}^{De}$}

\label{A2}

To acquire an analytical estimation of the ratio $%
P_{ave}^{De}/P_{ave}^{NonDe}$, we first simplify Eq. (\ref{1}) by assuming
the light intensity is unchanged in the vapor cell and neglecting the
derivative of the slow-down factor $q\left( P_{e}\right) $. Under these
approximations, we can express $P_{e}^{De}$ for depolarizing boundaries as $%
P_{e}^{De}=f_{xy}(x,y)f_{z}(z)$, where its longitudinal dependence is given
by:
\begin{equation}
f_{z}\left( z\right) =\frac{\tilde{R}_{op}}{\tilde{R}_{op}+\Gamma _{rel}}%
\left( 1-\frac{e^{-z/\lambda _{D}}+e^{-\left( L-z\right) /\lambda _{D}}}{%
1+e^{-L/\lambda _{D}}}\right) ,
\end{equation}%
with $\tilde{R}_{op}\equiv g_{P}\mathcal{L}\left( \Delta \right) I_{0}$.
This simplified solution shows that the depolarizing wall only affects a
layer with a width of $\lambda _{D}$. Though the solution for the radial
direction is more complicated, we can assume that the influence of the
depolarizing wall is similar. Hence, the \textquotedblleft
unfected\textquotedblright\ volume can be estimated as
\begin{equation}
\pi \left( R-\lambda _{D}\right) ^{2}\left( L-2\lambda _{D}\right) .
\end{equation}

Up until now, we have assumed an overall constant light intensity, $I_{0}$,
in the cell. However, the wall-depolarization length $\lambda_{D}$ should be
larger because $I$ decays in both the radial and longitudinal directions.
This decay is faster when the light is closer to the walls where the
polarization is small, and we can characterize it using the absorption
length $\lambda_{L}$. Accounting for this by including a factor of $\sqrt{%
1-\lambda_{D}/\lambda_{L}}$ in the ratio $P_{ave}^{De}/P_{ave}^{NonDe}$, we
obtain
\begin{eqnarray}
\frac{P_{ave}^{De}}{P_{ave}^{NonDe}} &\approx &\sqrt{1-\frac{\lambda _{D}}{%
\lambda _{L}}}\frac{\pi \left( R-\lambda _{D}\right) ^{2}\left( L-2\lambda
_{D}\right) }{\pi R^{2}L}  \notag \\
&=&\sqrt{1-\frac{\lambda _{D}}{\lambda _{L}}}\left( 1-2\frac{\lambda _{D}}{L}%
\right) \left( 1-\frac{\lambda _{D}}{R}\right) ^{2}.
\end{eqnarray}

For SERF magnetometers, if we ignore the spatial dependence of the light
intensity and the atomic polarization, the optimal transverse signal
\begin{equation*}
\tilde{P}_{x}=\frac{\gamma _{e}B}{2\left( R_{op}+\Gamma _{rel}\right) },
\end{equation*}%
while the $z$-direction polarization is $1/2$. Taking into account of the
wall-depolarizing effect, a factor $P_{ave}^{De}/P_{ave}^{NonDe}\ $is added
and the largest transverse signal $\tilde{P}_{x,ave}^{De}$ for depolarizing
walls becomes%
\begin{equation}
\tilde{P}_{x,ave}^{De}\approx \max \left[ \frac{P_{ave}^{De}}{P_{ave}^{NonDe}%
}\frac{\gamma _{e}B}{2\left( R_{op}^{0}+\Gamma _{rel}\right) }\right] .
\label{a2}
\end{equation}%
However, for large densities of $N_{2}$, Eq. (\ref{a2}) may overestimate the
maximal $P_{x,ave}^{De}$. Back to the full expression for the
spatial-independent transverse signal
\begin{equation}
P_{x}=\frac{\gamma _{e}BR_{op}}{\left( R_{op}+\Gamma _{rel}\right) ^{2}},
\end{equation}%
we obtain Eq. (\ref{7}).

%\bibliographystyle{plain}
%\bibliography{ref}

\bibliography{v2.bbl}

\end{document}